\documentclass[pre,twocolumn,showpacs]{revtex4}
\usepackage{epsfig}

\begin{document}

\title{Gaussian variational ansatz in the problem of anomalous sea waves: 
Comparison with direct numerical simulations}
\author{V. P. Ruban}
\email{ruban@itp.ac.ru}
\affiliation{Landau Institute for Theoretical Physics RAS, Moscow, Russia} 

\date{\today}

\begin{abstract}
The nonlinear dynamics of an obliquely oriented wave packet at sea surface is studied both 
analytically and numerically for various initial parameters of the packet, in connection 
with the problem of oceanic rogue waves. In the framework of Gaussian variational ansatz 
applied to the corresponding (1+2D) hyperbolic nonlinear Schr\"o\-dinger equation, a simplified 
Lagrangian system of differential equations is derived, which determines the evolution of 
coefficients of the real and imaginary quadratic forms appearing in the Gaussian. This model 
provides a semi-quantitative description for the process of nonlinear spatio-temporal focusing, 
which is one of the most probable mechanisms of rogue wave formation in random wave fields.
The system is integrated in quadratures, which fact allows us to understand qualitative 
differences between the linear and nonlinear regimes of the focusing of wave packet.
Comparison of the Gaussian model predictions with results of direct numerical simulation
of fully nonlinear long-crested water waves is carried out.

\vspace{3mm}

\end{abstract}


\maketitle

\section{Introduction}

Anomalous waves at the ocean surface and in many other physical systems (known also as
giant waves, rogue waves, freak waves) have been a popular subject of present-day 
scientific research (see, e.g., reviews \cite{Kharif-Pelinovsky,DKM2008,ORBMF-2013},
and references therein). The most frequently used mathematical model for this phenomenon is
the (1+1D) nonlinear Schr\"o\-dinger equation (NLSE) with the  focusing nonlinearity.
In particular, such equation describes the complex amplitude of the main harmonic of
a quasi-monochromatic weakly nonlinear surface wave over plane potential flows of an ideal 
fluid \cite{Zakharov68}. Due to nonlinear self-focusing, the modulation instability develops
in a sufficiently lengthy and high wave group \cite{Zakharov68,BF}. As the result, 
a one-dimensional anomalous wave arises. Such a scenario is confirmed by numerical 
and laboratory experiments \cite{ZDV2002,DZ2005Pisma,ZDP2006,R2012,CHA2011,CHOA2012}.
The full integrability makes the (1+1D) NLSE attractive for analytical studies and 
provides many exact solutions which describe some important properties of real
rogue waves (see, e.g., \cite{AEK85,AK86,AEK87,AAS-C2009,EHKFAD2011}).

In the Nature however sea waves are rather far from the one-dimensional model
(see, e.g., \cite{OOS2006,OWT2009,R2006PRE,R2007PRL,Peregrine,LP2006,ShS2014}).
When on the free fluid surface there are disturbances depending on both horizontal coordinates,
the wave dynamics becomes more complicated already in the framework of the corresponding
NLSE, not to speak about strongly nonlinear regimes. First, (1+2D) NLSE is not integrable.
Second, in the case of deep-water gravity waves, the spatial differential operator in it 
appears hyperbolic:
\begin{equation}
2i\psi_t+\psi_{xx}-\psi_{yy}+|\psi|^2\psi=0.
\label{NLS}
\end{equation}
Here the non-dimensional variables are in use: $k_0 A^*\rightarrow \psi$,
$\omega_0 t \rightarrow t$, $2k_0 x\rightarrow x$, $\sqrt{2}k_0 y\rightarrow y$,
where $A(x,y,t)$ is the complex envelope of the main harmonic, $k_0$ is the wave number, 
$\omega_0=2\pi/T_0=\sqrt{gk_0}$ is the frequency of the carrier wave
(the reference frame is moving with the group velocity 
$v_{\rm gr}=(1/2)\sqrt{g/k_0}$ along $x$ axis). 
Although Eq.(\ref{NLS}) has some  disadvantages, which are not discussed here,
but it is more suitable for approximate analytical studies (because in all its modified 
variants, some pseudo-differential operators appear inevitably, difficult to calculate them
``by hands'').

Thus, in the transverse direction the nonlinearity acts in the defocusing manner, while
in the longitudinal direction it works for focusing of a wave packet. In addition, the linear 
dispersion makes its own defocusing contribution. Owing to the indicated factors, 
the behavior of a nonlinear group of waves on the two-dimensional free surface of 
the fluid is more diverse as compared to the one-dimensional case. Depending on initial 
conditions, the nonlinearity can in some cases  reinforce the linear focusing, but as well 
it can destroy the focusing in other cases. In the case of reinforcement, a rogue wave arises. 
Besides that, anomalous waves can appear as a result of interaction between previously formed
coherent structures \cite{R2007PRL}. However, for the usual oceanic wave fields where the 
presence of coherent structures is hardly probable, more actual seems the scenario when 
anomalous waves rise due to  occasional spatio-temporal focusing (see \cite{FGD2007,R2013}, 
and references therein).

In general, the question about optimal conditions for the nonlinear focusing of water waves
is still far from being clear. It is only obvious that, similarly to the 1D case, the initial
wave group should be sufficiently high and/or extensive. But an effect of geometric shape
of the packet, as well as phase modulation, on the development of anomalous wave is poorly 
studied. In the absence of exact (1+2D) NLSE solutions, which could describe quantitatively
all such processes, it seems reasonable to carry out approximate consideration of the dynamics
of an idealized solitary wave packet characterized by a few time-dependent parameters.
Thereby, the above qualitative reasons could be concretized, and a reference point could be put
along the direction of development of future more accurate theory of three-dimensional rogue waves.
 
The purpose of this work is to investigate the nonlinear dynamics of a wave packet
in the framework of the full Gaussian variational ansatz, i.e. in the case when 
the quadratic form under the exponential has an off-diagonal part. Such a packet has an elliptical
shape, with main axes oriented at some time-dependent angle with respect to the coordinate axes.
Contrary to the more widely known diagonal ansatz (see, e.g., 
\cite{DAL1991,B1994,BR1996,BRKST1996,B1998,BEC1,BEC2,CD1996,H2002}), the full Gaussian ansatz
was previously applied to the elliptic NLSE only  \cite{DBDK2010,ADO2011}. The present work fills 
this gap in the theory. Here a system of variational equations will be derived for parameters
of the ellipse, and its full integrability will be demonstrated. The fact of integrability
takes place, first, because the third power of the NLSE nonlinearity is accorded with
the two spatial dimensions, and second, because  Eq.(\ref{NLS}) has a special integral of motion,
namely the hyperbolic analog of the angular momentum,
\begin{equation}
\int i[y(\psi_x\psi^*-\psi\psi^*_x)+x(\psi_y\psi^*-\psi\psi^*_y)]dx dy =const.
\end{equation}

The knowledge of general structure of analytical solutions of the variational model will allow us
to understand qualitative differences between the linear and nonlinear regimes of the wave 
packet focusing. Besides that, with the purpose of immediate testing of the Gaussian model, 
a comparison between its predictions and the results of numerical simulations of fully nonlinear
long-crested waves will be carried out. We shall see that the agreement can be rather good even
for strongly nonlinear regimes, when the very NLSE is not applicable already.

\section{Variational ansatz}

Let us first remind that in the simplest --- diagonal --- variant, the Gaussian ansatz describes
an elliptical wave packet having the symmetry axes coinciding with the coordinate axes
(see, e.g., \cite{DAL1991,B1994,BR1996,BRKST1996,B1998,BEC1}):
\begin{equation}
\psi=\sqrt{\frac{4N}{XY}}\exp\Big[-\frac{x^2}{2X^2}-\frac{y^2}{2Y^2} 
+i\frac{Ux^2}{2X} -i\frac{Vy^2}{2Y} +i\phi\Big].
\label{Gaussian}
\end{equation}
Besides the longitudinal size $X(t)$ and transverse size $Y(t)$, such a packet is characterized
by four more real quantities: $U$, $V$, $N$, and $\phi$. Parameter $N$ does not depend on time
because $4\pi N=\int |\psi|^2 dx dy$ is an exact integral of motion of NLSE, while the conjugate
variable $\phi$ is cyclic. Parameters $U$ and $V$ describe a phase modulation, with 
positive/neg\-ative values of $U$ or $V$ corresponding to defocusing/focusing  configurations
respectively in $x$ or $y$ direction. Here it is necessary to note that for applicability of NLSE,
the condition of narrow spectral width of the wave packet should be satisfied. In our case it
practically means the following: $(X,Y)\gtrsim 10$, $(U,V)\lesssim 0.1$. Equations of motion
determining the temporal behavior of the unknown  quantities $X$, $Y$, $U$, and $V$, are derived by 
substitution of the variational ansatz (\ref{Gaussian}) into the Lagrangian functional of the NLSE:
\begin{equation}
{\cal L}=\int(i\psi_t\psi^*-i\psi\psi^*_t-|\psi_x|^2+|\psi_y|^2+|\psi|^4/2) dx dy.
\label{L_NLSE}
\end{equation}
In this way we obtain a reduced Lagrangian $\tilde{\cal L}$ depending on 
$N$, $X(t)$, $Y(t)$, $U(t)$, and $V(t)$. The action integral $\int\tilde{\cal L}dt$ 
should be then variated. The main outcome of this standard procedure is the following.
Variation of the reduced action gives in particular that $U=\dot X$, $V=\dot Y$. 
Finally we arrive at the following system of differential equations of the Newtonian type:
\begin{equation}
\ddot X=\frac{1}{X^3}-\frac{N}{X^2Y}, \qquad \ddot Y=\frac{1}{Y^3}+\frac{N}{Y^2X}.
\label{XY_dyn}
\end{equation}
Note that equations (\ref{XY_dyn}) constitute a Lagrangian system with the Lagrangian 
functional $L\propto \tilde{\cal L}/N$, where
\begin{equation}
2L=\dot X^2-\dot Y^2 -\frac{1}{X^2} +\frac{1}{Y^2}+\frac{2N}{XY},
\label{L_XY}
\end{equation}
so the kinetic energy is not positively-definite (the second ``particle'' 
has a negative mass).

It should be said that equations of the type (\ref{XY_dyn}) and their generalizations
are actively used in plasma physics, in nonlinear optics, and in atomic physics to investigate
behavior of NLSE and Gross-Pitaevskii equation solutions in two and three spatial dimensions
(see papers \cite{DAL1991,B1994,BR1996,BRKST1996,B1998,BEC1,BEC2,CD1996,H2002}, 
and references therein). In particular in work \cite{B1994}, analytical and some numerical
solutions were presented namely for the hyperbolic (1+2D) NLSE, as in our case, but with such 
parameter values that are typical for nonlinear optics, not for water waves. 
In relation to the problem of anomalous sea wave focusing, this simplified variational ansatz
was applied only very recently in the author's work \cite{R2014}.

Let us now turn our attention to the full Gaussian ansatz, i.e. when off-diagonal elements 
of the quadratic forms are involved. For further consideration, it is convenient to introduce
linear combinations of the spatial coordinates, corresponding to a hyperbolic rotation
with the parameter $\chi$ (analogously to the elliptic NLSE case, but with the difference 
that there the rotation was usual trigonometric):
\begin{equation}
\bar x= x\cosh \chi+ y\sinh\chi,\qquad \bar y=x\sinh\chi +y\cosh\chi.
\end{equation}
Instead of expression (\ref{Gaussian}) we now write
\begin{equation}
\psi=\sqrt{\frac{4N}{XY}}\exp\Big[-\frac{\bar x^2}{2X^2}-\frac{\bar y^2}{2Y^2} 
+i\frac{U\bar x^2}{2X} -i\frac{V\bar y^2}{2Y} -i\gamma \bar x\bar y +i\phi\Big].
\label{Gaussian_full}
\end{equation}
Strictly speaking, with $\chi\not= 0$ the parameters  $X$ and $Y$ are no longer the 
longitudinal and transverse sizes of the packet, but we shall continue to call them so
for brevity. Parameter $\chi$ cannot be too large, otherwise the wave packet becomes
spectrally wide and goes beyond the NLSE applicability domain. Practically, as it will 
be seen from the results of direct numerical simulation of nonlinear 
waves with realistic parameters, the inequality $|\chi|\lesssim 0.5$ should take place.

Note that Gaussian form (\ref{Gaussian_full}) is one of exact solutions of the 
corresponding {\em linear} Schr\"odinger equation, with appropriate temporal dependencies
of the parameters appearing there. Consequently, at small $N$ the approximation  
(\ref{Gaussian_full}) is certainly valid. In practice, values $N\approx$ 2--4  
are of interest. In this case, generally speaking, only qualitative agreement
of the results of variational model (\ref{Gaussian_full}) with the solutions 
of the NLSE (and, the more so, with the completely 
nonlinear dynamics of waves on water) can be expected. In more detail the comparison
between the variational and numerical solutions will be discussed some later.
An undoubted benefit of this approximation is that it  provides a
semiquantitative description of the process of spatio-temporal focusing,
which is one of the most probable mechanisms of the formation of rogue waves 
under real conditions (see \cite{FGD2007,R2013}, and references therein). 

Having substituted the trial function (\ref{Gaussian_full}) and its partial derivatives
(expressed it terms of $\bar x$ and $\bar y$) into the Lagrangian (\ref{L_NLSE}), after
straightforward calculations we obtain that variable $\chi$ is cyclic, while the 
corresponding conserved quantity is (taking into account that $N=const$)
\begin{equation}
M=\gamma(X^2+Y^2)=const.
\end{equation}
Herewith, the temporal dependence of parameter $\chi$ is determined by the equation
\begin{equation}
\dot\chi=M(Y^2-X^2)/(X^2+Y^2)^2.
\label{d_chi}
\end{equation}
The conserved quantity $M$ is a hyperbolic analog of the angular momentum.
As previously, we have $U=\dot X$, $V=\dot Y$, but the dynamics of unknown functions
$X(t)$ and $Y(t)$ is described now by the following Lagrangian:
\begin{equation}
2L_M=\dot X^2-\dot Y^2 -\frac{1}{X^2} +\frac{1}{Y^2}+\frac{2N}{XY}
+\frac{M^2(X^2-Y^2)}{(X^2+Y^2)^2}.
\label{LM_XY}
\end{equation}
This Lagrangian is a hyperbolic analog of the elliptic variational problem considered
in works \cite{DBDK2010,ADO2011}. Equations of motion that follow from here, have the form 
\begin{equation}
\ddot X=\frac{1}{X^3}-\frac{N}{X^2Y} +\frac{M^2X}{(X^2+Y^2)^2}\Bigg[
1-2\frac{(X^2-Y^2)}{(X^2+Y^2)}\Bigg],
\label{ddX_M}
\end{equation}
\begin{equation}
\ddot Y=\frac{1}{Y^3}+\frac{N}{Y^2X} +\frac{M^2Y}{(X^2+Y^2)^2}\Bigg[
1-2\frac{(Y^2-X^2)}{(X^2+Y^2)}\Bigg].
\label{ddY_M}
\end{equation}
The finding solutions of this system and the comparison of the corresponding evolution of  
wave packet with a direct numerical modeling of nonlinear waves is the essence 
of the present work.

\section{Integrability of the model}

An important advantage of system (\ref{LM_XY}) is an exact integrability, because here 
in analogy with the elliptic case, the separation of variables turns out to be possible.
Integrability of the variational approximation is typical namely for NLSE in 2D space,
because in 3D the reduced Lagrangian instead of $2N/(XY)$ would contain $2\tilde N/(XYZ)$,
and it would destroy the homogeneity of the equations. It is well known from the analytical 
mechanics that for integration of the above system one should use the hyperbolic coordinates:
\begin{equation}
X=Q\sinh \xi, \qquad Y=Q\cosh \xi.
\end{equation}
These coordinates describe sector $X<Y$ which is most interesting in application to
the problem of oceanic rogue waves, because a length of their crest usually 3--10 times
exceeds the wave length $\lambda_0=2\pi/k_0$, while the longitudinal packet size
in the minimum reaches approximately one  $\lambda_0$  (that corresponds to dimensionless
values $X_{\rm min}\sim$ 6--10). The Lagrangian (\ref{LM_XY}) takes the form
\begin{equation}
2L_M=Q^2\dot \xi^2-\dot Q^2 -\frac{F(\xi)}{Q^2},
\label{L_Qxi}
\end{equation}
where function $F(\xi)$ is defined by the formula 
\begin{equation}
 F(\xi)=\frac{4}{\sinh^2(2\xi)}-\frac{4N}{\sinh(2\xi)}+\frac{M^2}{\cosh^2(2\xi)}.
\label{F}
\end{equation}
The corresponding equations of motion are 
\begin{eqnarray}
\ddot Q+Q\dot\xi^2+\frac{F(\xi)}{Q^3}&=&0,
\label{ddQ}
\\
2Q\dot Q\dot \xi +Q^2\ddot\xi +\frac{F'(\xi)}{2Q^2}&=&0.
\label{dd_xi}
\end{eqnarray}
After multiplying Eq.(\ref{dd_xi}) by $2Q^2\dot\xi$, we obtain in its left hand side a total 
time-derivative, so the integral of motion follows:
\begin{equation}
Q^4\dot\xi^2+F(\xi)=C=const.
\label{C}
\end{equation}
Looking at the expression (\ref{F}), it is easy to understand that at sufficiently small
negative values of the integration constant $C$, equation $F(\xi)=C$ determining 
``turn points'', has two real positive roots  ($\xi_1$ and $\xi_2$), with the motion 
of variable $\xi$ taking place between them. At  $C\ge 0$ the ``turn point'' is sole.

Eq.(\ref{ddQ}) is simplified to $\ddot Q=-C/Q^3$. Its general solution is written in explicit
form as 
\begin{equation}
Q(t)=\sqrt{I(t-t_*)^2-C/I},
\label{Q}
\end{equation}
where $t_*$ and $I$ are two more integration constants. It is easy to check that $I=-2E$, 
where $E$ is the energy integral of the system: 
\begin{equation}
E=(Q^2\dot \xi^2-\dot Q^2 +F(\xi)/Q^2)/2.
\end{equation}
With taking into account Eq.(\ref{Q}), it follows from Eq.(\ref{C}) that
\begin{equation}
\int_0^{t}\frac{dt}{I(t-t_*)^2-C/I}=\pm\int_{\xi_0}^{\xi}\frac{d\xi}{\sqrt{C-F(\xi)}},
\label{t_xi}
\end{equation}
where $\xi_0$ is the fourth --- the last --- integration constant. Formulas  (\ref{Q}) and 
(\ref{t_xi}) provide the full solution of the problem. In particular, with 
$I>0$, $C<0$, the dependence (\ref{t_xi}) can be represented in the following form,
\begin{equation}
\sinh 2\xi(t)=
P\Big(a_0+\frac{1}{\sqrt{-C}}\arctan\Big[\frac{I(t-t_*)}{\sqrt{-C}}\Big]\Big),
\label{sh2xi}
\end{equation}
where an even $a$-periodic function $P(a)$ (having a minimum at  $a=0$) is the inverse
with respect to elliptic integral (here the integration variable $z=\sinh 2\xi$):
\begin{equation}
P\Big(\pm\int_{s_1}^s\frac{z dz}{2\sqrt{(1+z^2)(Cz^2+4Nz-4)-M^2z^2}}\Big)\equiv s,
\label{P_def}
\end{equation}
and $s_1$ is the smaller root of the equation $(1+s^2)(Cs^2+4Ns-4)-M^2s^2=0$.
The amplitude of wave packet equals to  $\sqrt{4N/XY}$, and it is given by the formula 
\begin{equation}
A(t)=\sqrt{8N}/\sqrt{[I(t-t_*)^2 -C/I]\sinh 2\xi(t)}.
\end{equation}

\section{Selection of the solutions}

\begin{figure}
\begin{center}
 \epsfig{file=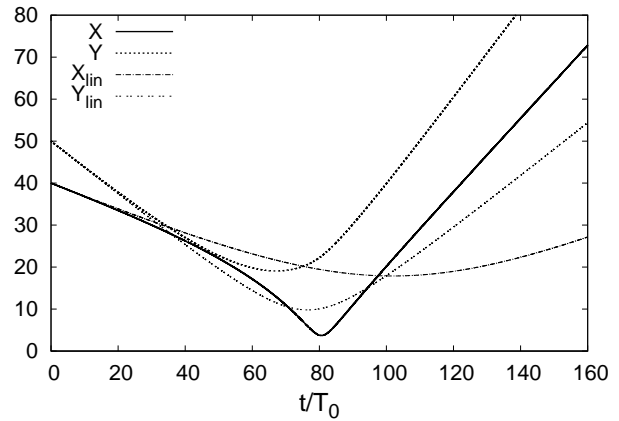,width=86mm}
\end{center}
\caption{The temporal dependencies $X(t)$ and $Y(t)$ for $N=3.0$, $M=0$, $X_0=40$, $Y_0=50$, 
$U_0=-0.05$, $V_0=-0.10$. The solutions for a linear packet with the same initial conditions 
are also shown for comparison.  What is important for formation of a rogue wave, the nonlinear 
focusing along the longitudinal direction goes here with acceleration for most of the time, 
while in the linear case it occurs with deceleration for all the time. The value of $X$ in the 
minimum has been too small for validity of the NLSE, but they are initial conditions of this type
which in reality result in formation of anomalous waves of the limiting steepness. 
} 
\label{N3XYt} 
\end{figure}

In fact, we shall not use the analytical solutions, expressed in terms of special functions,
for finding the dependencies  $X(t)$ and $Y(t)$ with some given initial values. Practically 
it is much simpler and faster to carry out highly accurate numerical simulation of the 
variational equations  (\ref{ddX_M}) and (\ref{ddY_M}) immediately. One should also remember
that the absolute accuracy in solving approximate variational problem is hardly necessary.
The knowledge of the analytical structure of the solutions will be needed for a general
understan\-ding of properties of the dynamical system under consideration. 

We are interested mainly in those solutions from the whole diversity (\ref{Q}) and (\ref{t_xi}), 
which result in essentially higher wave amplitude as compared to the maximally possible linear value 
$A_{\rm max\, lin}=\sqrt{4N U_{-\infty} V_{-\infty}}$, where $U_{-\infty}$ and $V_{-\infty}$
are the asymptotic values of the focusing parameters at large negative times. 
In essence, only such events are deserving the terrible title ``rogue waves''.
What is important, the most interesting solutions correspond to values of the parameters 
$C\approx 0$, $M\approx 0$. The condition  $C=0$ means that $Q^2\equiv Y^2-X^2=I(t-t_*)^2$,
i. e. the plots of functions $X(t)$ and $Y(t)$ touch each other at $t=t_*$.
It is easy to make sure of the above assertion by numerical simulations of 
Eqs.(\ref{ddX_M})-(\ref{ddY_M}) [with fixed $N$, $U_0$, $V_0$, $Y_0$, but different 
$X_0$ and $M$], and subsequent comparison of the maximal amplitudes. Typical dependencies
$X(t)$ and $Y(t)$ for such a case are shown in Fig.\ref{N3XYt}. Let us also note that
the minimal value of the ratio $(X/Y)_{\rm min}=\tanh\xi_1$, which is reached closely to
the moment of maximal amplitude, at $M=0$ and $C=0$ can be found from the simple condition
$\sinh 2\xi_1=1/N$, as it is clear in view of formula (\ref{P_def}).

\begin{figure}
\begin{center}
\epsfig{file=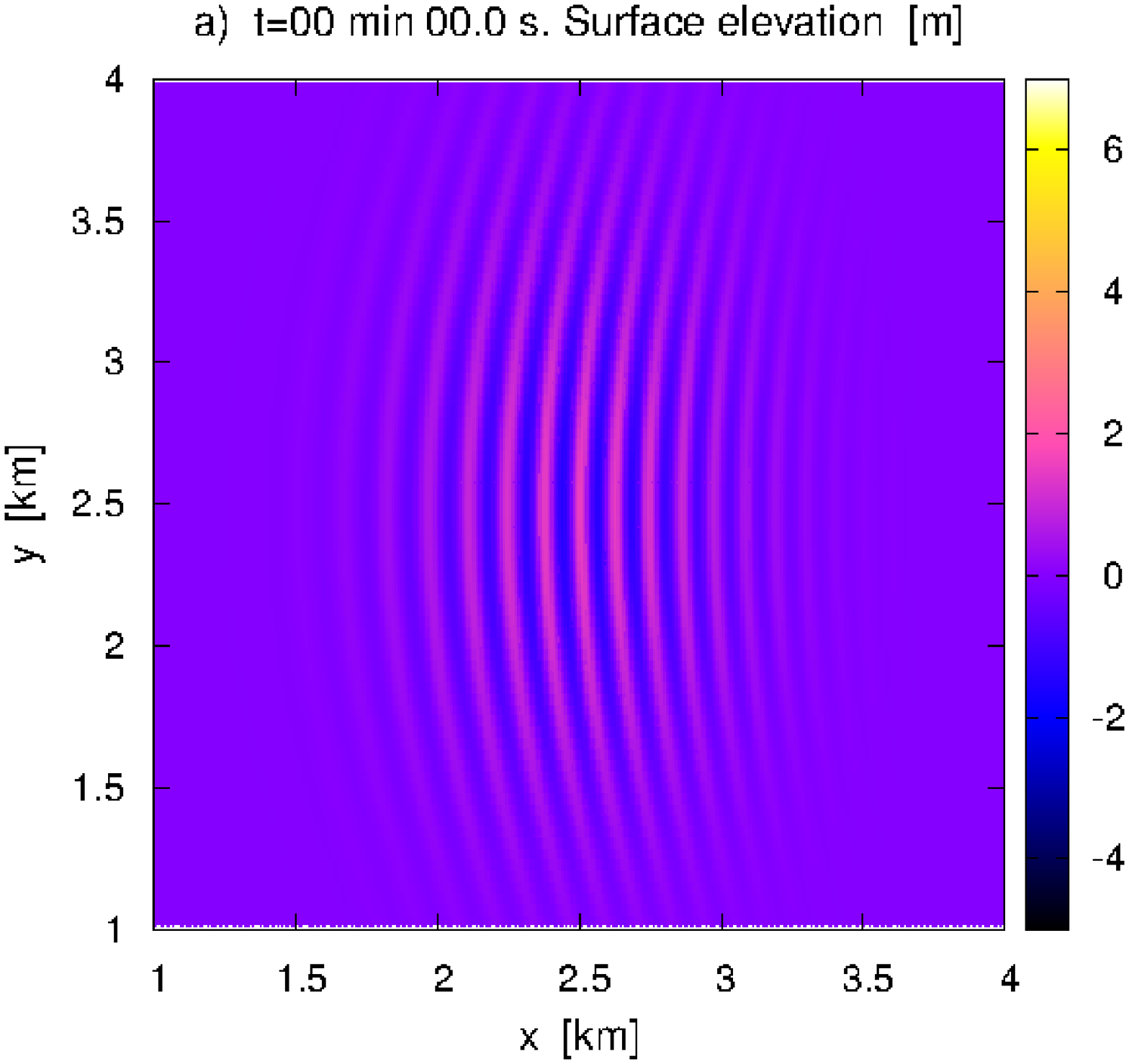,width=82mm}\\
\vspace{4mm}
\epsfig{file=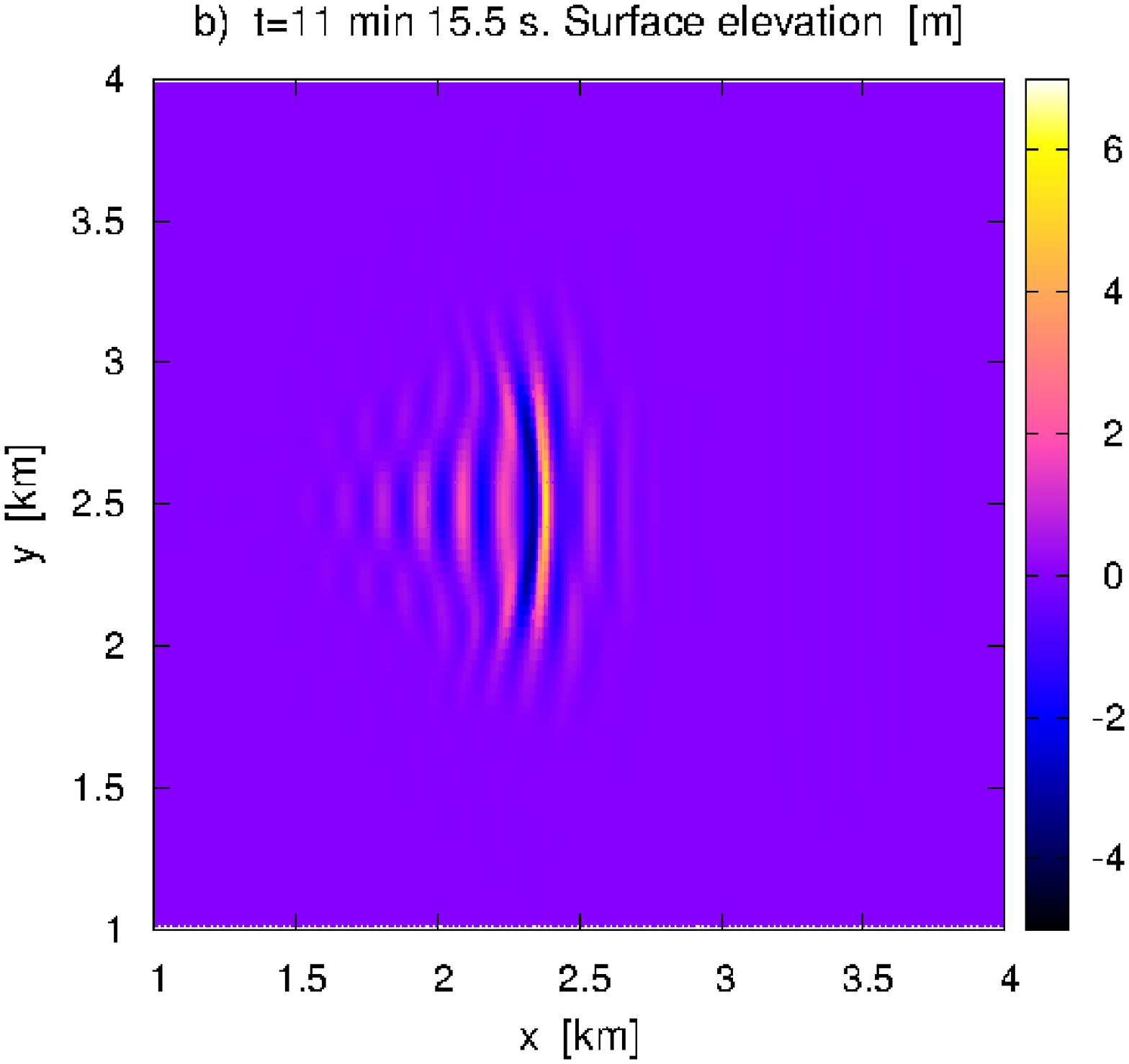,width=82mm}
\end{center}
\caption{Wave picture at при $\chi=0.0$ (the other parameters are the same as in Fig.1):
a) wave packet at the initial time moment; b) at the moment when the anomalous wave has risen.
The crest length of the high wave here has been much longer than it would be in the linear case,
which observation is in accordance with the Gaussian model prediction. Deviations from the 
Gaussianity are clearly seen, basically --- in the form of a small tail consisting of longer waves.} 
\label{N3chi00XY} 
\end{figure}

\begin{figure}
\begin{center}
   \epsfig{file=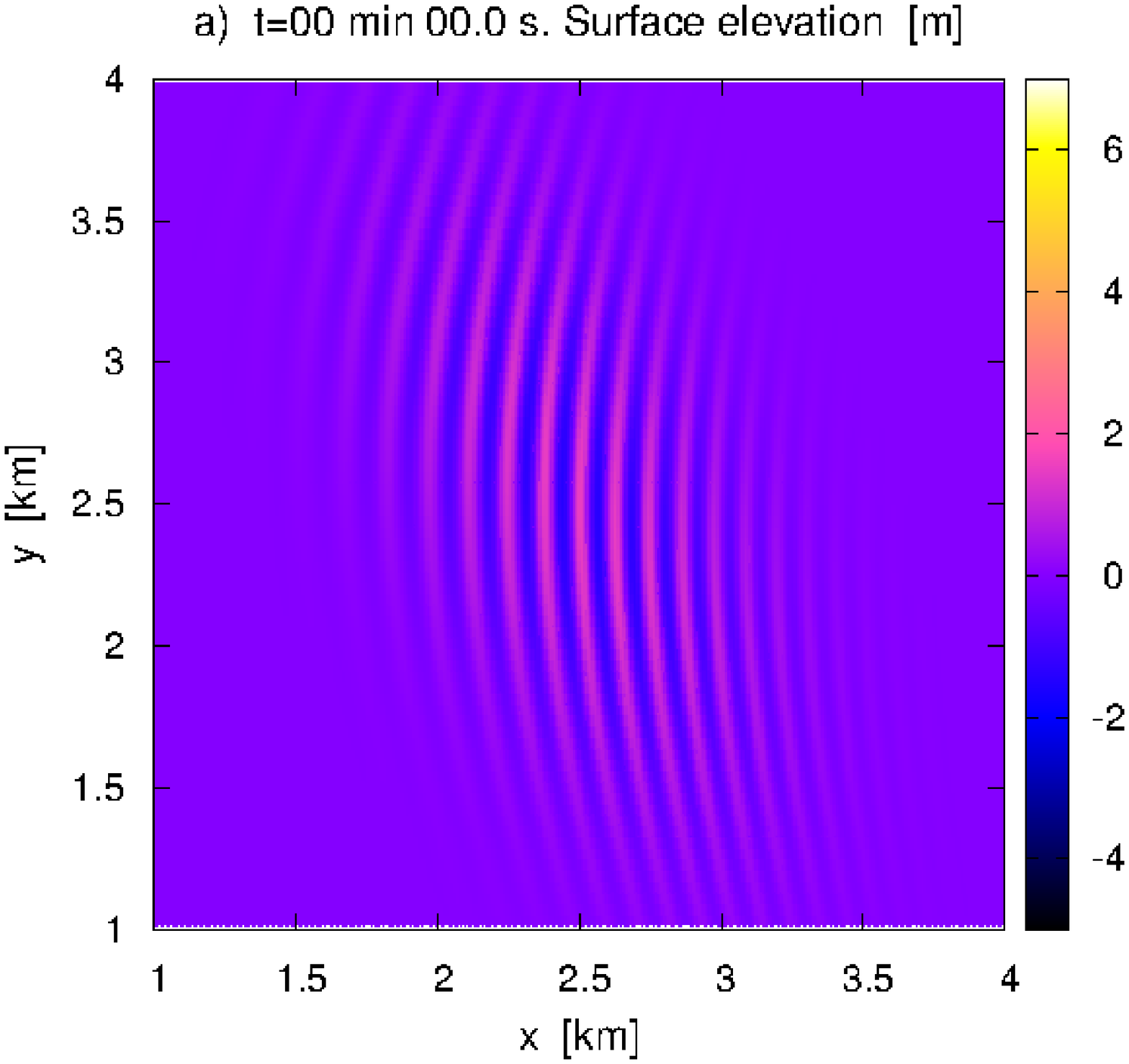,width=82mm}\\
\vspace{4mm}
   \epsfig{file=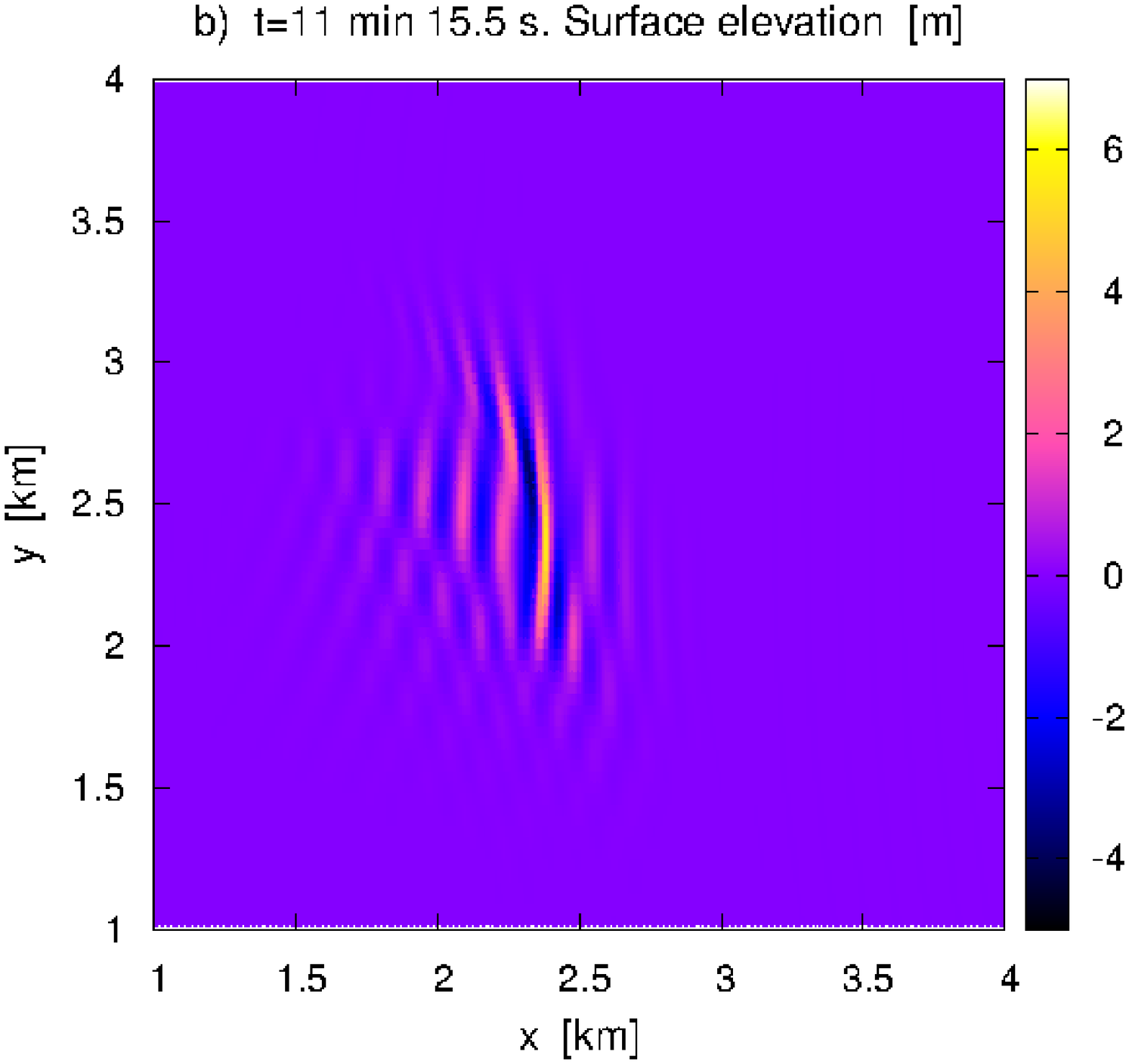,width=82mm}
\end{center}
\caption{Wave picture at при $\chi=0.3$ (the other parameters are the same as in Fig.1):
a) wave packet at the initial time moment; b) at the moment of the anomalous wave rise.
Deviations from the Gaussianity are even more prominent here than at   $\chi=0.0$.} 
\label{N3chi03XY} 
\end{figure}

It is worth noting that at $M=0$ the cyclic variable $\chi$ remains constant, as it follows
from Eq.(\ref{d_chi}). As we shall see soon, the physical wave fields corresponding to
different values of $\chi$, strongly differ between each other. In particular, at $\chi=0$
the wave picture is symmetric with respect to sign change of $y$ coordinate, while
at $\chi\not =0$ there is no any symmetry (see Figs.\ref{N3chi00XY}-\ref{N3chi03XY}).
But in the framework of NLSE, the envelope of one solution coincides with the envelope of the
other solution after a hyperbolic rotation of the dimensionless coordinates. The presence
of this approximate symmetry significantly clarifies the question about optimal conditions
for the formation of anomalous waves.

\begin{figure}
\begin{center}
   \epsfig{file=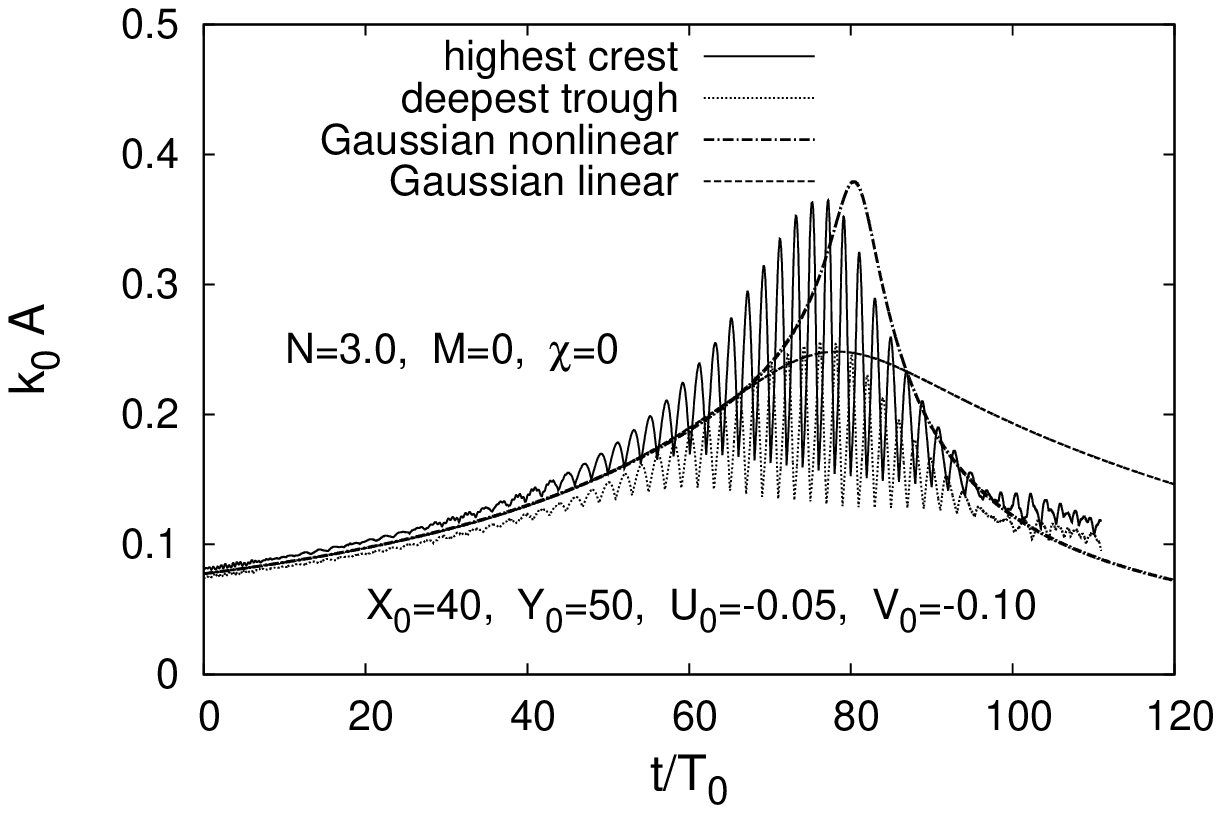,width=85mm}
   \epsfig{file=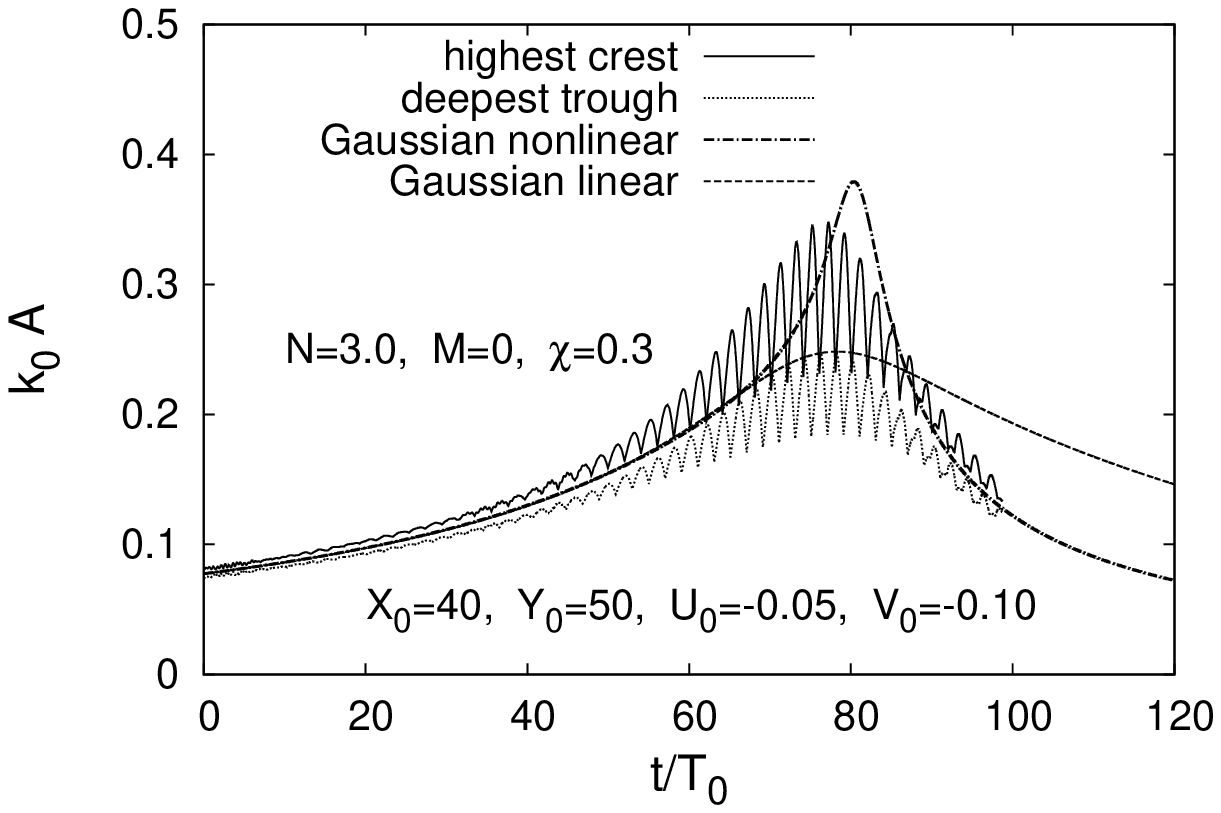,width=85mm}\\
   \epsfig{file=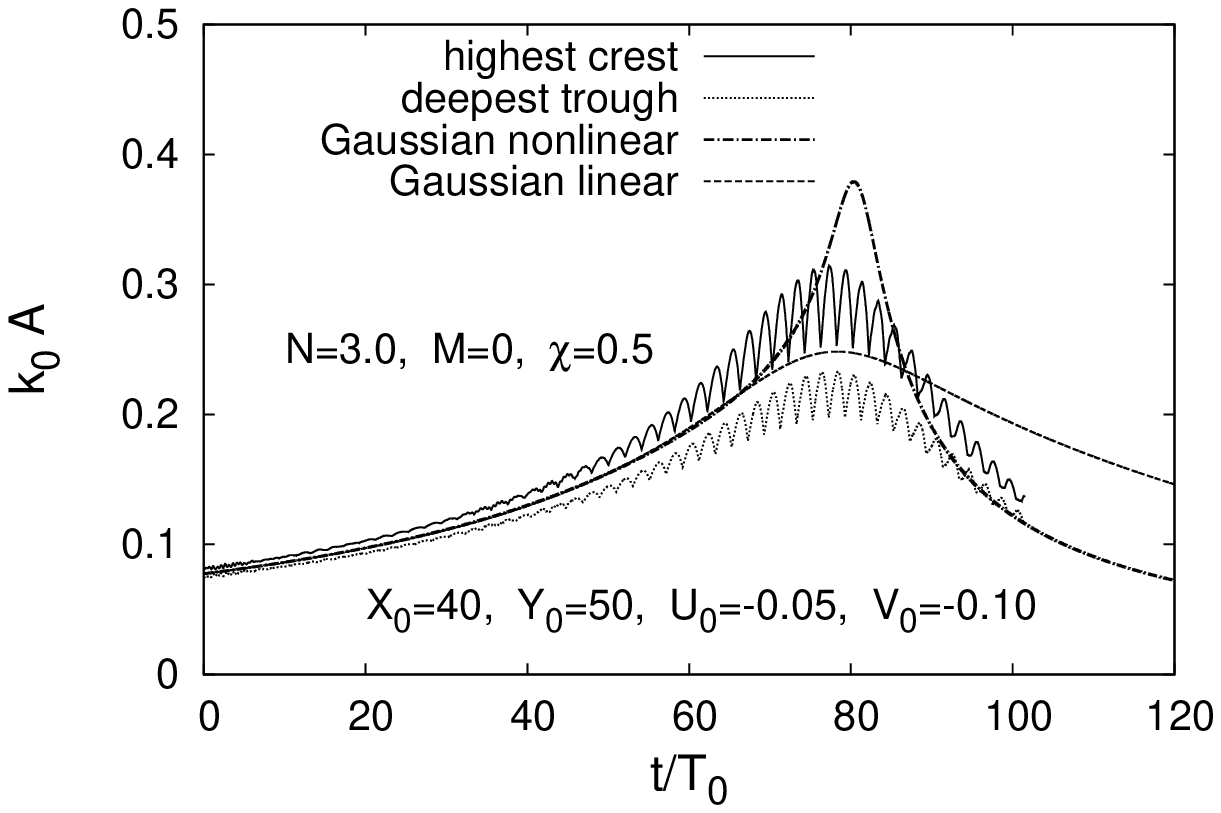,width=85mm}
\end{center}
\caption{Comparison of the temporal dependencies of the maximal amplitudes for wave packets
with different $\chi$,  the other parameters being the same as in Fig.1. 
In each figure shown are numerically found the height of the highest crest and the depth
of the deepest trough, and also predictions of the Gaussian models --- linear and nonlinear.
The numerical curves have an oscillating character, which property is related to the 
approximately two-fold difference between the phase velocity of the carrier wave and the group 
velocity of the packet motion. Heights of crests are typically larger than depths of 
troughs, due to the presence of higher harmonics. Therefore the quantity that should be 
compared to the theoretical predictions is a half-sum of envelopes of the numerical curves.} 
\label{N3kA} 
\end{figure}

\begin{figure}
\begin{center}
   \epsfig{file=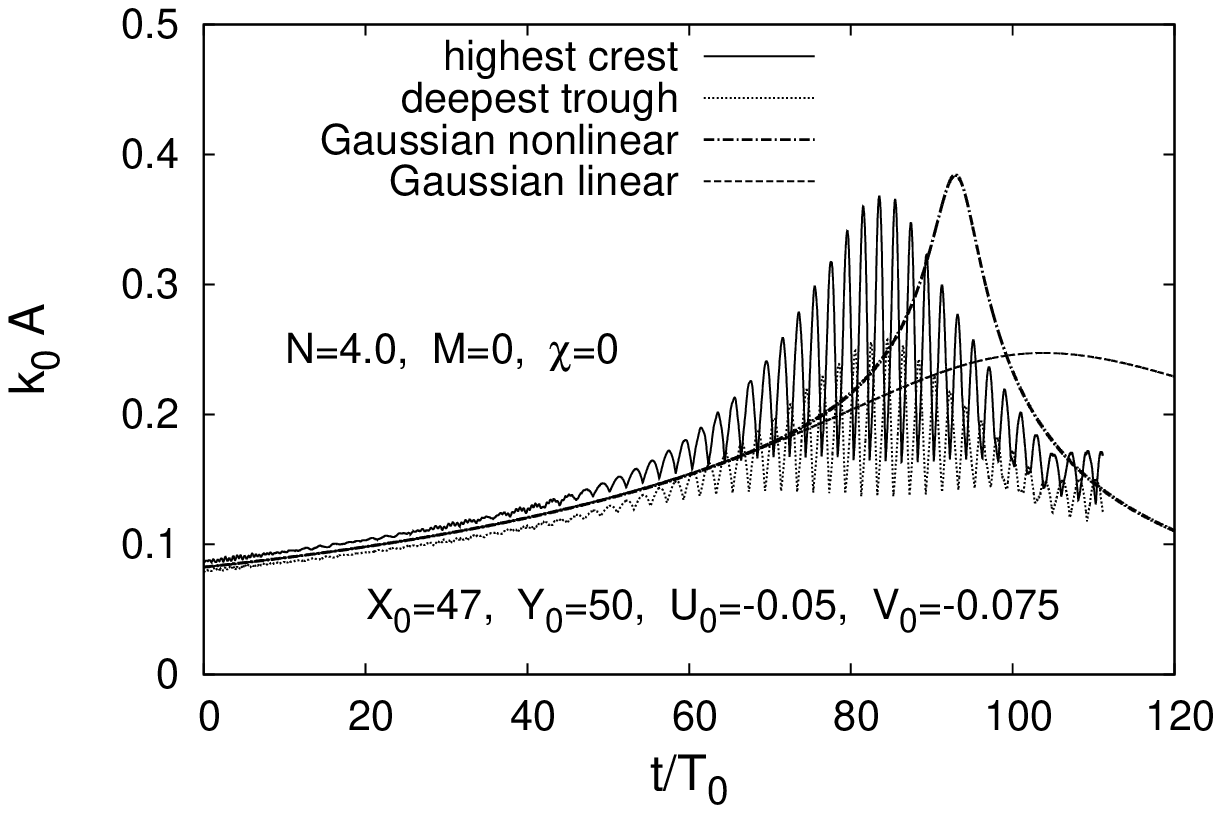,width=85mm}
   \epsfig{file=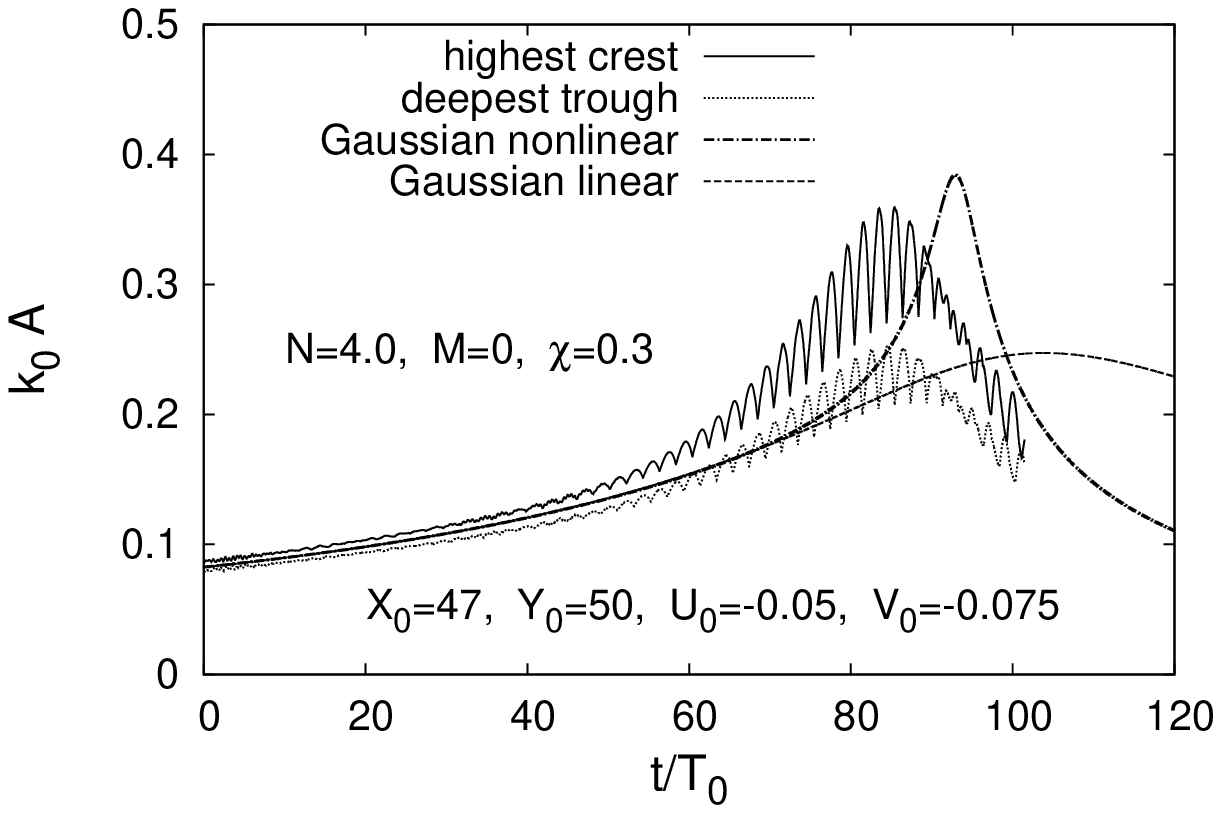,width=85mm}
\end{center}
\caption{Comparison of the maximal amplitudes for different sets of initial data 
than in Fig.4. It is seen that with increase of $N$, the time delay of the predicted peak 
in comparison with the numerical simulation increases.} 
\label{N4kA} 
\end{figure}

\section{Comparison with numerical experiments}

It is clear that any simplified model without verification remains just a toy.
In order to make the status of the variational approximation more distinct, it is necessary to
compare it with a more accurate model. In our case, numerical experiments were carried out
in the framework of the fully nonlinear  long-crested wave model which is described in detail
in Refs.\cite{RD2005,R2010}. The computational domain was a square of the size 5 km, with
the periodic boundary conditions. The length of the carrier wave was chosen  $\lambda_0=125$ m.
The simulations were carried out for different sets of initial parameters of the packet. 

In Fig.2a and Fig.3a, examples are shown of how Gaussian initial conditions look. 
The ``portraits'' of anomalous waves developing from those conditions, are presented in
Fig.2b and Fig.3b. In general, deviations from the Gaussian shape are clearly seen there.
Despite the non-Gaussianity of the arising rogue waves, Figs.4-5 demonstrate that there is
a qualitative agreement between the variational theory and the numerical experiment in the  
important aspect as the temporal dependence of the maximal amplitude of the wave packet.  
It is seen that Gaussian model somewhat overestimates the height of anomalous wave
(especially as parameter $\chi$ increases to values about unity, when the approximate 
symmetry of the hyperbolic rotation becomes invalid), and it moves to the future the moment 
of the maximal rise. However, it is hardly reasonable to expect a detailed accordance
from the simplest variational approximation. In any case, the time period of existence
of anomalous wave is predicted by the Gaussian theory rather well. Also another important 
effect is confirmed, the increase of the transverse packet size as compared to the 
linear theory (see Fig.1). This effect contributes to the subsequent more fast decrease
of the amplitude in comparison with the linear case. 

\section{Conclusions}

Thus, based on the variational analysis and on the comparison with the results
of direct numerical simulations, it is natural to put forward the following thesis.
In the main approximation, a typical long-crested three-dimensional oceanic rogue wave 
at the nonlinear stage is, from the mathematical point of view, a dynamical system with 
just two degrees of freedom: $X(t)$ and  $Y(t)$. This system contains two constant parameters
($N$ and $M$), which are related to the integrals of motion of 1+2D NLSE and characterize 
a given group of waves. The system is integrable. Real sea states where one can expect 
anomalous waves, require values $N\approx$ 2--4. The most favorable focusing parameters
for the rogue wave formation are: 1) $M\approx 0$; 2) on the initial stage of (occasional)
focusing at some time moment  $t_*$ the relations $X_*\approx Y_*\sim 20-40$, 
$-\dot X_*\approx -\dot Y_* \gtrsim$ 0.05 should take place. To estimate probability of
realization of such conditions in random weakly nonlinear wave fields is the task on the future.

\end{document}